\documentclass[10pt]{article}
\usepackage{latexsym}
\usepackage{amssymb}
\usepackage{amsmath}
\usepackage{amscd}
\usepackage{amsthm}
\usepackage[left=2cm,top=2.5cm,right=2.5cm,bottom=1.5cm]{geometry}
\usepackage[dvips]{graphicx}
\usepackage{epstopdf}
\usepackage{hyperref}
\begin{document}
\begin{center}
\large{\bf{Interacting Viscous Dark Energy in Bianchi Type-III Universe}} \\
\vspace{10mm}
\normalsize{Hassan Amirhashchi} \\
\vspace{5mm}
\normalsize{Young Researchers and Elite Club, Mahshahr Branch, Islamic Azad University, Mahshahr, Iran \\
\vspace{2mm}
E-mail: h.amirhashchi@mahriau.ac.ir;~~~hashchi@yahoo.com} \\
\end{center}
\vspace{10mm}
\begin{abstract}
In this paper we study the evolution of the equation of state of
viscous dark energy in the scope of Bianchi type III space-time.
We consider the case when the dark energy is minimally coupled to
the perfect fluid as well as direct interaction with it. The
viscosity and the interaction between the two fluids are
parameterized by constants $\zeta_{0}$ and $\sigma$ respectively.
We have made a detailed investigation on the cosmological
implications of this parametrization. To differentiate between
different dark energy models, we have performed a geometrical
diagnostic by using the statefinder pair $\{s, r\}$.
\end{abstract}

\section{Introduction}
Recent Astronomical and astrophysical observations indicate that we live in an accelerating expanding universe
(Perlmutter et al. 1997, 1999; Riess et al. 1998, 2001; Tonry et al. 2003; Tegmark et al. 2004). This fact opens a very fundamental
question regarding to the source which can produce such an accelerating expansion. Since the ordinary matter (energy)
generates an attractive gravitational force, there should be a kind of un-known, non-baryonic source of energy with negative
pressure in order to make the expansion of the universe to be accelerating. Of course, the amount of this energy should be
larger than the ordinary matter (energy) since first a fraction of this force has to counterbalance the attractive force of ordinary
matter and then the rest give rise to acceleration. According to the recent observations we live in a nearly spatially flat
Universe composed of approximately $4\%$ baryonic matter, $22\%$ dark matter and $74\%$ dark energy (DE). We know that the ultimate
fate of our universe will be determined by dark energy but unfortunately our knowledge about its nature and properties is
still very limited. It is not even known what is the current value of the dark energy effective equation of state (EoS) parameter
$\omega^{X} = p^{X}/\rho^{X}$. We only know that a kind of exotic energy with negative pressure drives the current accelerating
expansion of the universe; and although it dominates the present universe, it was small at early times. This is why so far many
candidates have been proposed for dark energy including: cosmological constant ($\omega^{X}=-1$) (Weinberg 1989; Carroll 2001; Padmanabhan 2003; Peebles \& Ratra 2003), quintessence ($-1<\omega^{X}<-\frac{1}{3}$) (Wetterich 1988; Ratra \& Peebles 1988), phantom ($\omega^{X}<-1$) (Caldwell 2002), quintom
($\omega^{X}<-\frac{1}{3}$) (Feng et al. 2005), interacting dark energy models, Chaplygin gas as well as generalized Chaplygin gas models
(Srivastava 2005; Bertolami et al. 2004; Bento et al. 2002; Alam et al. 2003), and etc. A cosmological constant (or vacuum energy) seems to be a proper candidate for dark energy which can explain the current acceleration in a natural way, but it would suffer from some theoretical problems such as the fine-tuning and coincidence problems. Quintessence and phantom dark energy models are provided by scalar fields. These models are also encounter to
some problems. For example, since recent observations (Hinshaw et al. 2009; Komatsu et al. 2009; Copland et al. 2006; Perivolaropoulos 2006) indicate that $\omega^{X}<-1$ is allowed at $68\%$ confidence level, quintessence with $\omega^{X}>-1$ may not be a proper candidate as dark energy. Phantom dark energy models are also suffer from some fundamental problems, such as future singularity problem called Big Rip (Caldwell ey al. 2003; Nesseris \& Perivolaropoulos 2004) and the ultraviolet quantum instabilities problem (Carroll et al. 2003). Since recent cosmological observations mildly favor models with a transition from $\omega^{X}>-1$ to $\omega^{X}<-1$ near the past (Riess et al. 2004; Choudhury \& Padmanabhan 2005), a combination of quintessence and phantom in a unified model called quintom has been proposed (Feng et al. 2005).\\

Recently the dissipative DE models in which the negative pressure, responsible for the current acceleration, is an
effective bulk viscous pressure have been proposed in order to avoid the occurrence of the big rip (McInnes 2002; Barrow 2004). The
general theory of dissipation in relativistic imperfect fluid was first suggested by Eckart (1940), Landau and Lifshitz
(1987). Although this is only the first-order deviation from equilibrium and may suffer from causality problem, one can still
apply it to phenomena which are quasi-stationary, i.e. slowly varying on space and time characterized by the mean free path and
the mean collision time. It is worth to mention that the second-order causal theory was obtained by Israel (1976) and
developed by Israel and Stewart (1976). The effect of bulk viscosity on the background expansion of the universe has been
investigated from different points of view (Cataldo et al. 2005; Bervik \& Gorbunova 2005; Szydlowski \& Hrycyna 2007; Singh 2008; Feng \& Zhou 2009; Oliver et al. 2011; Amirhashchi 2013a,b). There are also some astrophysical observational evidences indicate
that the cosmic media is not a perfect fluid (Jaffe et al. 2005). Therefore, the viscosity effect could be concerned in the evolution of the
universe. The role of viscous pressure as an agent that drives the present acceleration of the Universe has also been studied in Refs
(Zimdhal et al. 2001; Balakin et al. 2003). The possibility of a viscosity dominated late epoch of the Universe with accelerated expansion was already
mentioned by Padmanabhan and Chitre (1987).\\

Interaction between dark energy and dark matter (DM) is a proposal suggested as a possible solution to the coincidence problem (Setare 2007; Jamil \& Rashid 2008, 2009; Cimento et al. 2003). Moreover, DE-DM interaction provides the possibility of detecting the dark energy in a natural way. It is worth to mention
that the possibility of such an interaction has been supported by the recent observations (Bertolami et al. 2007; Le Delliou et al. 2007; Berger \& Shojaei 2006). Interacting dark energy models have been widely investigated in literatures (for example see Amirhashchi et al. 2011 a, b; Amirhashchi et al. 2012; Amirhashchi et al. 2013 ; Amirhashchi 2013a,b,c; Saha et al. 2012; Yadav and Sharma 2013; Yadav 2012; Pradhan et al 2011; Setare 2007a,b,c; Setare et al 2009; Sheykhi \& Setare 2010; Jamil \& farooq 2010; Zhang 2005; Sajadi \& Vodood 2008 ). A Full dynamical analysis of anisotropic scalar-field cosmology with arbitrary potentials has been studied by Fadragas et al (2013). Recently, Long Zu et al. (2014) have investigated a class of transient acceleration models consistent with Big Bang Cosmology. In this paper, we study the behavior of the viscous dark energy EoS parameter in an anisotropic space-time namely Bianchi type III universe in the following two cases: (i) when DE and DM are minimally coupled i.e there is no any interaction between these two dark components and (ii) when there is an interaction between viscous DE and DM. We parameterize the interaction by a constant $\sigma$ and viscosity by $\zeta_{0}$, then a detailed investigation of the cosmological implications of this parametrization will be provided by assuming an energy flow from DE to DM. Finally, to discriminate the different interaction parameters, as usual, a statefinder diagnostic is also performed.
\section{The Metric and Field  Equations}
We consider the Bianchi type-III metric as
\begin{equation}
\label{eq1}
ds^{2} = -dt^{2} + A^{2}(t)dx^{2}+B^{2}(t)e^{-2\alpha x}dy^{2}+C^{2}(t)dz^{2},
\end{equation}
where $A(t), B(t)$ and $C(t)$ are functions of time only. \\

We define the following physical and geometric parameters to be used in formulating the law and further in solving
the Einstein's field equations for the metric (\ref{eq1}). \\

The average scale factor $a$ of Bianchi type-III model (\ref{eq1}) is defined as
\begin{equation}
\label{eq2} a = (ABC)^{\frac{1}{3}}.
\end{equation}
A volume scale factor V is given by
\begin{equation}
\label{eq3} V = a^{3} = ABC.
\end{equation}
We define the generalized mean Hubble's parameter $\rm H$ as
\begin{equation}
\label{eq4} H = \frac{1}{3}(H_{x} + H_{y} + H_{z}),
\end{equation}
where $\rm H_{x} = \frac{\dot{A}}{A}$, $\rm H_{y} = \frac{\dot{B}}{B}$ and $\rm H_{z} = \frac{\dot{C}}{C}$ are the
directional Hubble's parameters in the directions of $x$, $y$ and $z$ respectively. A dot stands for differentiation
with respect to cosmic time $t$. \\

From Eqs. (\ref{eq2})-(\ref{eq4}), we obtain
\begin{equation}
\label{eq5} H = \frac{1}{3}\frac{\dot{V}}{V} = \frac{\dot{a}}{a} = \frac{1}{3}\left(\frac{\dot{A}}{A} +
\frac{\dot{B}}{B} + \frac{\dot{C}}{C}\right).
\end{equation}
The physical quantities of observational interest in cosmology i.e. the expansion scalar $\theta$, the average
anisotropy parameter $Am$ and the shear scalar $\sigma^{2}$ are defined as
\begin{equation}
\label{eq6} \theta = u^{i}_{;i} = \left(\frac{\dot{A}}{A} + \frac{\dot{B}}{B} + \frac{\dot{C}}{C} \right),
\end{equation}
\begin{equation}
\label{eq7}\sigma^{2} =
\frac{1}{2}\left(\sum_{i=1}^{3}H_{i}^{2}-\frac{1}{3}\theta^{2}\right),
\end{equation}
\begin{equation}
\label{eq8} A_{m} = \frac{1}{3}\sum_{i = 1}^{3}{\left(\frac{\triangle H_{i}}{H}\right)^{2}},
\end{equation}
where $\triangle H_{i} = H_{i} - H (i = x, y, z)$ represents the directional Hubble parameter in the direction of
$x$, $y$, $z$ respectively. $A_{m} = 0$ corresponds to isotropic expansion. \\

The Einstein's field equations ( in gravitational units $8\pi G = c = 1 $) read as
\begin{equation}
\label{eq9} R^{i}_{j} - \frac{1}{2} R g^{i}_{j} = - T^{(m)i}_{j} -
T^{(X)i}_{j},
\end{equation}
where $T^{mi}_{j}$ and $T^{Xi}_{j}$ are the energy momentum
tensors of perfect fluid and viscous DE, respectively. These are
given by
\[
  T^{(m)i}_{j} = \mbox{diag}[-\rho^{m}, p^{m}, p^{m}, p^{m}],
\]
\begin{equation}
\label{eq10} ~ ~ ~ ~ ~ ~ ~ ~  = \mbox{diag}[-1, \omega^{m},
\omega^{m}, \omega^{m}]\rho^{m},
\end{equation}
and
\[
 T^{(X)i}_{j} = \mbox{diag}[-\rho^{X},~ p^{X}, ~p^{X},~ p^{X}],
\]
\begin{equation}
\label{eq11} ~ ~ ~ ~ ~ ~ ~ ~ ~ ~ ~ ~ ~ ~ = \mbox{diag}[-1,
\omega^{X},~ \omega^{X},~ \omega^{X}]\rho^{X},
\end{equation}
where $\rho^{m}$ and $p^{m}$ are, respectively the energy density
and pressure of the perfect fluid component or ordinary baryonic
matter while $\omega^{m} = p^{m}/\rho{m}$ is its EoS parameter.
Similarly, $\rho^{X}$ and $p^{X}$ are, respectively the energy
density and effective pressure of the DE component while
$\omega^{X}= p^{X}/\rho^{X}$
is the corresponding EoS parameter. \\
In Eckart's theory (1940) a viscous dark energy EoS is
specified by
\begin{equation}
\label{eq12} {p}^{X}_{eff} = p^{X}+ \Pi.
\end{equation}
Here $\Pi = -\xi(\rho^{X})u^{i}_{;i}$ is the viscous pressure and
$H = \frac{u^{i}_{;i}}{3}$ is the Hubble's parameter. On
thermodynamical grounds, in conventional physics $\xi$ has to be
positive. This is a consequence of the positive sign of the
entropy change in an irreversible process (Nojiri \& Odintsov 2003). In
general, $\xi(\rho^{X})=\xi_{0}(\rho^{X})^{\tau}$, where
$\xi_{0}>0$ and $\tau$ are constant parameters.\\

In a co-moving coordinate system ($u^{i} = \delta^{i}_{0}$),
Einstein's field equations (\ref{eq9}) with (\ref{eq10}) and
(\ref{eq11}) for Bianchi type-III metric (\ref{eq1}) subsequently
lead to the following system of equations:
\begin{equation}
\label{eq13} \frac{\ddot{B}}{B} + \frac{\ddot{C}}{C} +
\frac{\dot{B}\dot{C}}{BC} = -\omega^{m}\rho^{m} -
\omega^{X}_{eff}\rho^{X}+\Pi,
\end{equation}
\begin{equation}
\label{eq14} \frac{\ddot{C}}{C} + \frac{\ddot{A}}{A} +
\frac{\dot{C}\dot{A}}{CA} = -\omega^{m}\rho^{m} -
\omega^{X}_{eff}\rho^{X}+\Pi,
\end{equation}
\begin{equation}
\label{eq15} \frac{\ddot{A}}{A} + \frac{\ddot{B}}{B} +
\frac{\dot{A} \dot{B}}{AB} - \frac{\alpha^{2}} {A^{2}} =
-\omega^{m}\rho^{m} - \omega^{X}_{eff}\rho^{X}+\Pi,
\end{equation}
\begin{equation}
\label{eq16} \frac{\dot{A}\dot{B}}{AB} + \frac{\dot{A}\dot{C}}{AC}
+ \frac{\dot{B}\dot{C}}{BC} - \frac{\alpha^{2}}{A^{2}} = \rho^{m}
+ \rho^{X},
\end{equation}
\begin{equation}
\label{eq17} \alpha\left(\frac{\dot{A}}{A} -
\frac{\dot{B}}{B}\right)  = 0.
\end{equation}
The law of energy-conservation equation ($T^{ij}_{;j} = 0$) yields
\begin{equation}
\label{eq18} \dot{\rho}^{m} + 3(1 + \omega^{m})\rho^{m}H +
\dot{\rho}^{X} +3(1 + \omega^{X}_{eff})\rho^{X}H = 0.
\end{equation}
The Raychaudhuri equation for given distribution is found to be
\begin{equation}
\label{eq19} \frac{\ddot{a}}{a} = \frac{1}{2}\xi\theta -
\frac{1}{6}(\rho^{X} + 3p^{X}) - \frac{1}{6}(\rho^{m} + 3p^{m}) -
\frac{2}{3}\sigma^{2}.
\end{equation}
\section{Solution of the Field Equations}
The field equations (\ref{eq13})-(\ref{eq17}) are a system of five
linearly independent equations with seven unknown parameters $A$,
$B$, $C$, $\rho^{m}$, $p^{X}$, $\rho^{X}$, $\omega^{X}$. Two
additional constraints relating
these parameters are required to obtain explicit solutions of the system.\\\\
Eq. (\ref{eq17}), obviously leads to
\begin{equation}
\label{eq20} B = \ell_{0} A,
\end{equation}
where $\ell_{0}$ is an integrating constant. \\

Firstly, we assume that the scalar expansion $\theta$ in the model
is proportional to the shear scalar. This assumption is in accord
with the Thorne study (Thorne 1967) which quotes that the observations of
the velocity red shift relation for extragalactic sources suggests
that Hubble expansion of the universe is isotropic today to
approximately within $30$ percent (Kantowski \& Sachs 1966; Kristian \& Sachs 1966; Mohanty et al. 2007). More precisely, red
shift studies place the limit $\frac{\sigma}{H}\leq 0.3$.
Therefore, from eqs. (\ref{eq5})-(\ref{eq7}) and (\ref{eq20}) we
get
\begin{equation}
\label{eq21} A = C^{n},
\end{equation}
where $n$ is a constant. \\

Secondly, following Amirhashchi et al (2011) we consider the
following ansatz for the scale factor
\begin{equation}
\label{eq22} a(t)=\sinh(t).
\end{equation}
By assuming a time varying deceleration parameter one can generate
such a scale factor. It has also been shown that this
scale factor is stable under metric perturbation (Chen et al. 2001). In
term of red shift the above scale factor turns to
\begin{equation}
\label{eq23} a=\frac{1}{1+z},~~~z=\frac{1}{\sinh(t)}-1.
\end{equation}

Now, by using (\ref{eq13}), (\ref{eq14}),
(\ref{eq20})-(\ref{eq23}) we can find the metric components as
\begin{equation}
\label{eq24} A = \ell_{1} \sinh^{\frac{3n}{2n +
1}}(t)=\ell_{1}(1+z)^{-\frac{3n}{2n + 1}},
\end{equation}
\begin{equation}
\label{eq25} B =\ell_{2} \sinh^{\frac{3n}{2n + 1}}(t)=\ell_{2}
(1+z)^{-\frac{3n}{2n + 1}},
\end{equation}
\begin{equation}
\label{eq26} C =\ell_{3} \sinh^{\frac{n}{2n + 1}}(t)=\ell_{3}
(1+z)^{-\frac{3}{2n + 1}},
\end{equation}
where $\ell_{1} = K^{-\frac{3n}{(2n + 1)}}$, $\ell_{2} = \ell_{0}\ell_{1}$, $\ell_{3} = \ell_{1}^{\frac{1}{n}}$ and
$K$ is an integrating constant. \\

Therefore, the metric (\ref{eq1}) reduces to
\[
ds^{2} = -dt^{2} + \ell_{1}^{2} \sinh^{\frac{6n}{2n+1}}(t) dx^{2}
+ \ell_{2}^{2} \sinh^{\frac{6n}{2n + 1}}(t) e^{-2\alpha x}dy^{2}
\]
\begin{equation}
\label{eq27} \ +\ell_{3}^{2} \sinh^{\frac{6}{2n+1}}(t)dz^{2}.
\end{equation}
One can write the above metric in terms of red shift as
\[
ds^{2} = -dt^{2} + \ell_{1}^{2} (1+z)^{-\frac{6n}{2n+1}} dx^{2} +
\ell_{2}^{2} (1+z)^{-\frac{6n}{2n + 1}} e^{-2\alpha x}dy^{2}
\]
\begin{equation}
\label{eq28} \ +\ell_{3}^{2} (1+z)^{-\frac{6}{2n+1}}(t)dz^{2}.
\end{equation}
In the following sections we deal with two cases, (i) viscous
non-interacting two-fluid model and (ii) viscous interacting two-
fluid model.
\section{Viscous Dark Energy (Non-Interacting Case)}
In this section we assume that two-fluid do not interact with each
other. Therefore, the general form of conservation equation
(\ref{eq18}) leads us to write the conservation equation for the
barotropic and dark fluid separately as,
\begin{equation}
\label{eq29}\dot{\rho}^{m} + 3\frac{\dot{a}}{a}\left(\rho^{m} +
p^{(m)}\right) = \dot{\rho}^{m} + (1 + \omega^{m})\rho^{m}(2n +
1)\frac{\dot{C}}{C} = 0,
\end{equation}
and
\begin{equation}
\label{eq30}\dot{\rho}^{X} + 3\frac{\dot{a}}{a}\left(\rho^{X} +
p^{X}_{eff}\right) = \dot{\rho}^{X} + (1 +
\omega^{X}_{eff})\rho^{X}(2n + 1)\frac{\dot{C}}{C} = 0.
\end{equation}
Integration of (\ref{eq29}) leads to
\begin{equation}
\label{eq31}\rho^{m} = \rho_{0}C^{-(2n + 1)(1 + \omega^{m})} =
\rho_{0}l_{0}\sinh^{-3(1 + \omega^{m})}(t)=\rho_{0}l_{0}(1+z)^{3(1
+ \omega^{m})},
\end{equation}
where $\rho_{0}$ is an integrating constant and $l_{0} = \ell_{3}^{-(2n + 1)(1 + \omega^{m})}$. \\

By using Eqs. (\ref{eq20}), (\ref{eq21}) and (\ref{eq31}) in Eqs.
(\ref{eq16}) and (\ref{eq13}), we obtain
\begin{equation}
\label{eq32} \rho^{X} = n(n + 2)\frac{\dot{C}^{2}}{C^{2}} -
\frac{\alpha^{2}}{C^{2n}} - \rho_{0}l_{0}\sinh^{-3(1 +
\omega^{m})}{(t)},
\end{equation}
\begin{figure}[ht]
\begin{minipage}[b]{0.5\linewidth}
\centering
\includegraphics[width=\textwidth]{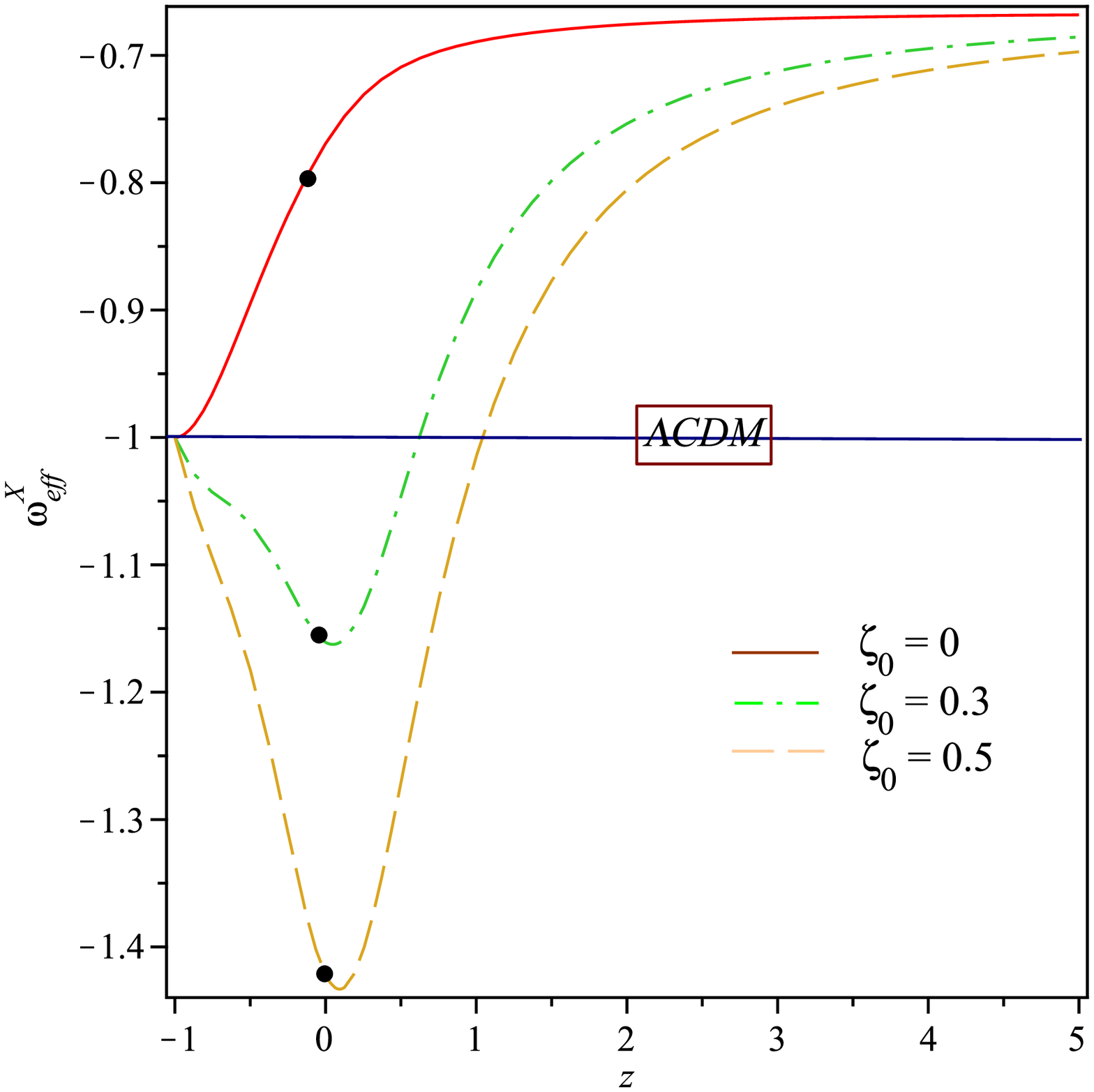} \\
\caption{The EoS parameter $\omega^{X}_{eff}$ versus $z$ for $n =
\beta = \alpha = \ell_{3} = l_{0}=1$, $\Omega_{0}^{m} = 0.3$. The
dots locate the current values of $\omega^{X}_{eff}$.}
\end{minipage}
\hspace{0.5cm}
\begin{minipage}[b]{0.5\linewidth}
\centering
\includegraphics[width=\textwidth]{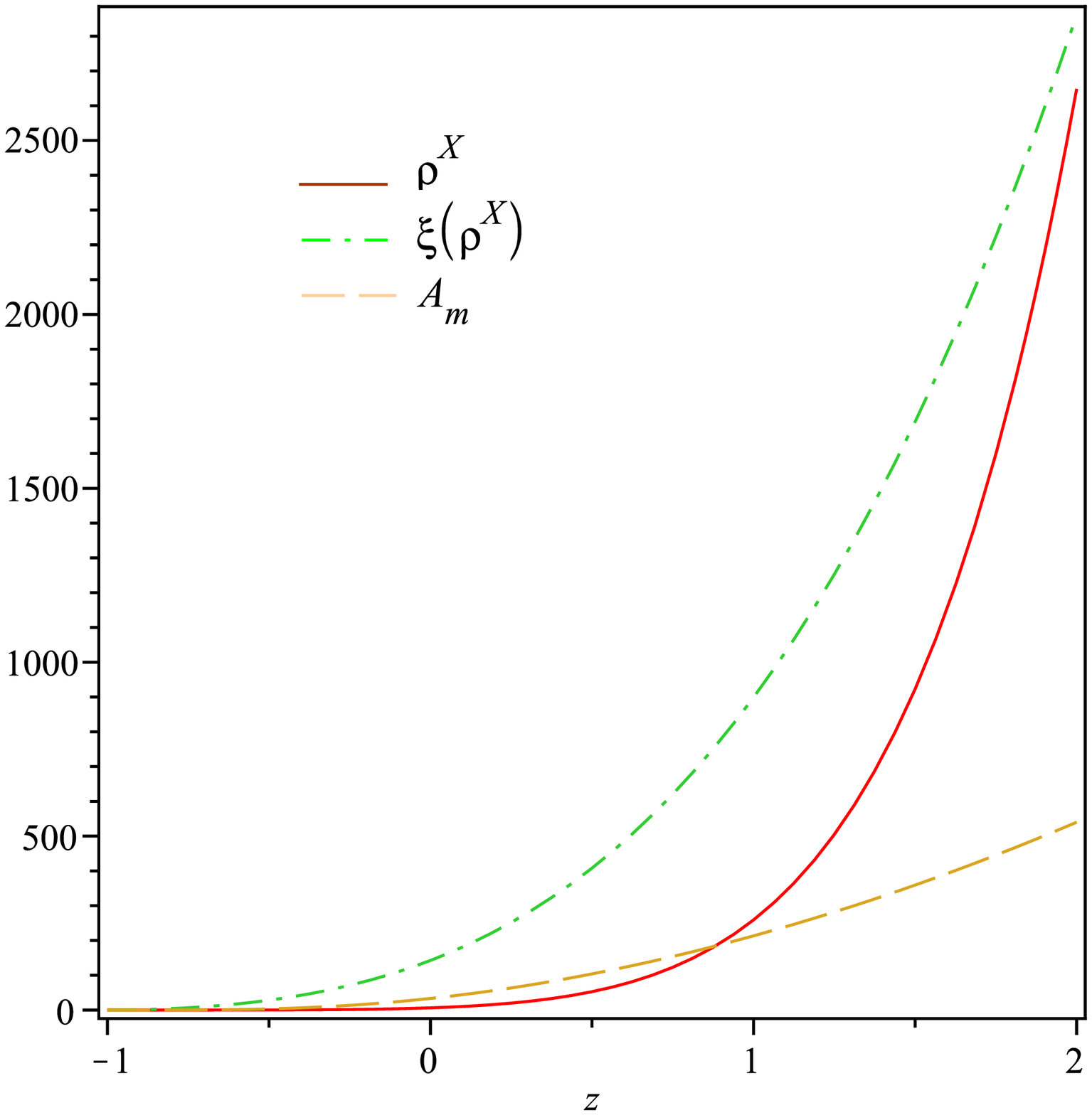}
\caption{The plot of the DE energy density $\rho^{X}$, average
anisotropy parameter $A_{m}$, and the bulk viscosity
$\xi(\rho^{X})$ vs. $z$ for $\alpha = \ell_{3} = \ell_{0}=1$, $
\Omega^{m}=0.3$, $\xi_{0}=0.1$.}
\end{minipage}
\end{figure}
and
\begin{equation}
\label{eq33} p^{X}= - \left[2n\frac{\ddot{C}}{C} +
n(3n-2)\frac{\dot{C}^{2}}{C^{2}} -
\frac{\alpha^{2}}{C^{2n}}\right] -
\omega^{m}\rho_{0}l_{0}\sinh^{-3(1 + \omega^{m})}(t).
\end{equation}
Using Eq. (\ref{eq23}) in Eqs. (\ref{eq32}) and (\ref{eq33}), we
obtain the energy density and pressure of DE i.e $\rho^{X}$ and
$p^{X}_{eff}$ as
\[
\rho^{X} = \frac{9n(n + 2)}{(2n + 1)^{2}}\coth^{2}(t) -
\alpha^{2}\ell^{-2n}_{3}\sinh^{-\frac{6n} {(2n + 1)}}(t) -
\rho_{0}l_{0}\sinh^{-3(1 + \omega^{m})}{(t)}
\]
\begin{equation}
\label{eq34}= \frac{9n(n + 2)}{(2n +
1)^{2}}\left[1+(1+z)^{2}\right] -
\alpha^{2}\ell^{-2n}_{3}(1+z)^{\frac{6n} {(2n + 1)}} -
\rho_{0}l_{0}(1+z)^{3(1 + \omega^{m})}
\end{equation}
\[
p^{X}_{eff}= -\left[\frac{9(n^{2} + n + 1)}{(2n +
1)^{2}}\coth^{2}(t)- \frac{3(n + 1)}{(2n + 1)} \cosh^{2}(t)\right]
- \omega^{m}\rho_{0}l_{0}\sinh^{-3(1 +
\omega^{m})}(t)-3\xi_{0}H(\rho^{X})^{\tau}
\]
\begin{equation}
\label{eq35}= -\left[\frac{9(n^{2} + n + 1)}{(2n +
1)^{2}}\left[1+(1+z)^{2}\right]- \frac{3(n + 1)}{(2n + 1)}
\left[1+(1+z)^{-2}\right]\right] -
\omega^{m}\rho_{0}l_{0}(1+z)^{3(1 +
\omega^{m})}-3\xi_{0}H(\rho^{X})^{\tau}.
\end{equation}
respectively. \\
Using above two equations we finally find the effective EoS
parameter of DE as
\[
\omega^{X}_{eff} = - \left[\frac{\frac{9(n^{2} + n + 1)}{(2n +
1)^{2}}\coth^{2}(t)- \frac{3(n + 1)}{(2n + 1)}\cosh^{2}{(t)} +
3\Omega^{m}_{0}l_{0}\omega^{m}\sinh^{-3(1 - \omega^{m})}(t)}
{\frac{9n(n + 2)}{(2n + 1)^{2}}\coth^{2}(t) -
\alpha^{2}\ell^{-2n}_{3} \sinh^{-\frac{6n}{(2n + 1)}}{(t)} -
3\Omega^{m}_{0}l_{0}\sinh^{-3(1 +
\omega^{m})}(t)}\right]-\zeta_{0}H^{\tau}(\Omega^{X})^{\tau-1}
\]
\begin{equation}
\label{eq36}= - \left[\frac{\frac{9(n^{2} + n + 1)}{(2n +
1)^{2}}\left[1+(1+z)^{2}\right]- \frac{3(n + 1)}{(2n +
1)}\left[1+(1+z)^{-2}\right] +
3\Omega^{m}_{0}l_{0}\omega^{m}(1+z)^{3(1 - \omega^{m})}}
{\frac{9n(n + 2)}{(2n + 1)^{2}}\left[1+(1+z)^{2}\right] -
\alpha^{2}\ell^{-2n}_{3} (1+z)^{\frac{6n}{(2n + 1)}} -
3\Omega^{m}_{0}l_{0}(1+z)^{3(1 +
\omega^{m})}}\right]-\zeta_{0}H^{\tau}(\Omega^{X})^{\tau-1}.
\end{equation}
Here $\zeta_{0}=3^{\tau}\xi_{0}$, $\Omega^{m}$, and $\Omega^{X}$
are the energy density of matter and DE respectively (note that
the subscript $0$ indicates the present value of any parameter).\\
The behavior of EoS parameter for dark energy in terms of red
shift $z$ is shown in Fig. $1$. Since we are interested in the
late time and future evolution of DE, we plot the range of red
shift $z$ from $-1$ to $z=5$. The parameter $\Omega^{m}$ is taken
to be $0$. This figure shows that the $\omega^{X}_{eff}$ of
non-viscous DE ($\xi_{0}=0$) is only varying in quintessence
region whereas the variation of viscose DE starts from
quintessence region, crossing PDL, and varies in phantom region.
But the EoS of both non-viscous and viscous DE ultimately
approaches to cosmological constant region ($\omega^{X}_{eff} =
-1$) independent of the value of $\xi_{0}$. This behavior clearly
shows that the phantom phase i.e $\omega^{X}_{eff}<-1$ is an
unstable phase and there is a transition from phantom to the
cosmological constant phase at late time. The variations of energy
density of $\rho^{X}$, mean anisotropy parameter $A_{m}$, and bulk
viscosity $\xi(\rho^{X})$ are depicted in Fig. $2$. As it is
expected all these parameters are decreasing functions and
approaches to zero at late tim ($z=-1$)
\\\\
The matter density $\Omega^{m}$ and dark energy density
$\Omega^{X}$ are also given by
\begin{figure}[ht]
\begin{minipage}[b]{0.5\linewidth}
\centering
\includegraphics[width=\textwidth]{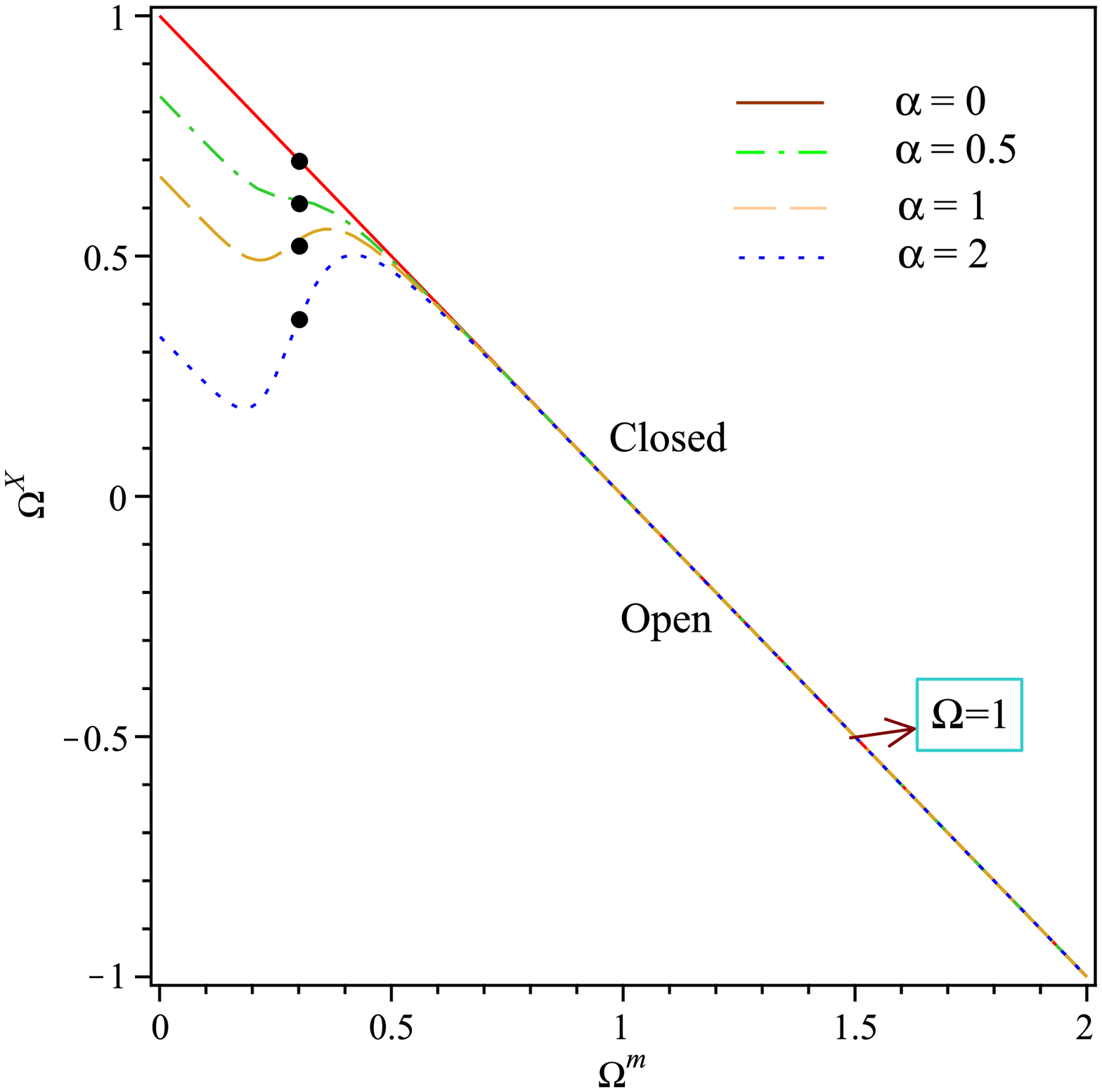} \\
\caption{The plot of $\Omega^{X}$ versus $\Omega^{m}$ for
$n=\ell_{0}=\ell_{3}=1,~\Omega^{m}_{0}=0.3$. The solid line
indicates flat universe ($n=1$,~$\alpha=0$). The dots locate the
current values of $\Omega^{X}$ and $\Omega^{m}$.}
\end{minipage}
\hspace{0.5cm}
\begin{minipage}[b]{0.5\linewidth}
\centering
\includegraphics[width=\textwidth]{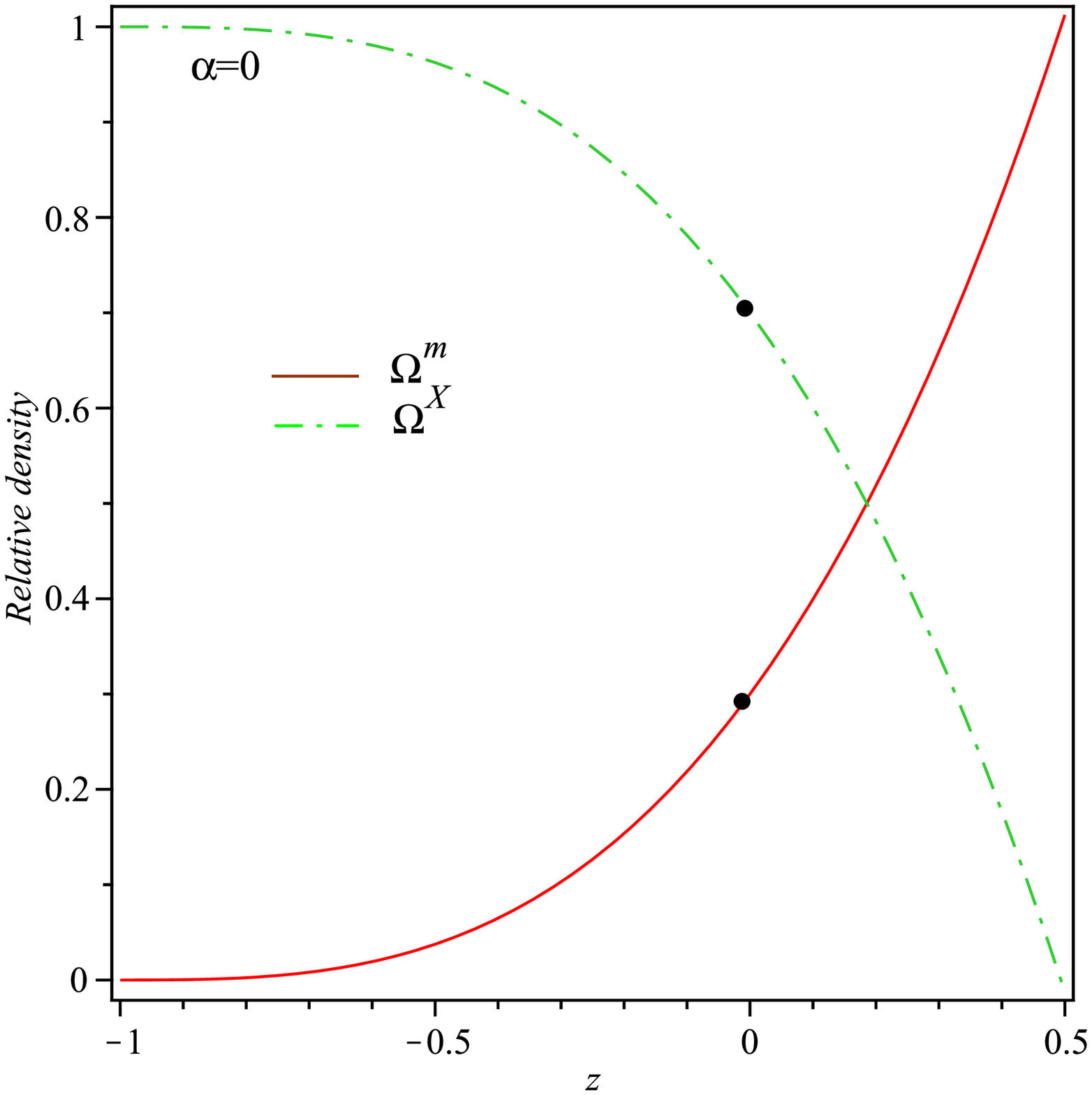}
\caption{ The plot of energy $\Omega^{m}$ and $\Omega^{X}$ versus
redshift ($z$) for $\Omega^{m}_{0}=0.3,~ \ell_{0}=n=1$. The dots
locate the current values of $\Omega^{X}$ and $\Omega^{m}$.}
\end{minipage}
\end{figure}
\begin{equation}
\label{eq37}\Omega^{m} = \frac{\rho^{m}}{3H^{2}} =
\frac{\rho_{0}l_{0}\sinh^{-3(1 - \omega^{m})}{(t)}}
{3\coth^{2}{(t)}}=\Omega^{m}_{0}l_{0}(1+z)^{3(1 - \omega^{m})},
\end{equation}
and
\[
\Omega^{X} =  \frac{\rho^{X}}{3H^{2}} =\frac{3n(n + 2)}{(2n +
1)^{2}}- \frac{\alpha^{2}\ell^{-2n}_{3}\sinh^{-\frac{6n}{(2n +
1)}}(t) + \rho_{0}l_{0}\sinh^{-3(1 + \omega^{m})}{(t)}}
{3\coth^{2}{(t)}}
\]
\begin{equation}
\label{eq38} =\frac{3n(n + 2)}{(2n + 1)^{2}}-
\frac{\alpha^{2}\ell^{-2n}_{3}(1+z)^{\frac{6n}{(2n + 1)}}}
{3\left[1+(1+z)^{2}\right]}-\Omega^{m}_{0}l_{0}(1+z)^{3(1 -
\omega^{m})}
\end{equation}
respectively. Adding Eqs. (\ref{eq37}) and (\ref{eq38}), we obtain
total energy ($\Omega$)
\[
\Omega = \Omega^{m} + \Omega^{X} =\frac{9n(n + 2)}{(2n +
1)^{2}}-\frac{\alpha^{2}\ell_{3}^{-2n}\sinh^{-\frac{6n}{2n +
1}}{(t)}}{3\coth^{2}{(t)}}
\]
\begin{equation}
\label{eq39}=\frac{9n(n + 2)}{(2n +
1)^{2}}-\frac{\alpha^{2}\ell_{3}^{-2n}(1+z)^{\frac{6n}{2n +
1}}}{3\left[1+(1+z)^{2}\right]}.
\end{equation}
Figure $3$ shows the values of $\Omega^{X}_{0}$ and
$\Omega^{m}_{0}$ which are permitted by our model. The line
$1=\Omega^{X}+\Omega^{m}$ represents a flat universe separating
open from closed universes. From this figure we observe that for
$\alpha=0,~n=1$ which represents a spatially flat universe
($\Omega=1$), $\Omega^{X}_{0}\approx 0.76$, and
$\Omega^{m}_{0}\approx 0.24$. Other models with different values
of $\alpha\neq 0$, represent various open
universes ($\Omega<1$).\\
The variation of density parameters $\Omega^{m}$ and $\Omega^{X}$
with red shift $z$ have been depicted in Fig. $4$. It is observed
that $\Omega^{X}$ increases as red shift decreases and approaches
to $1$ at late time whereas
$\Omega^{m}$ decreases as $z$ decreases and approaches to zero at late time.\\

\section{Viscous Dark Energy (Interacting Case)}
In this section we consider the interaction between dark and
barotropic fluids. For this purpose we can write the continuity
equations for barotropic and dark fluids as
\begin{equation}
\label{eq40}\dot{\rho}^{m} + 3\frac{\dot{a}}{a}\left(\rho^{m} +
p^{m}\right) = \dot{\rho}^{m} + (1 + \omega^{m}) \rho^{m}(2n +
1)\frac{\dot{C}}{C} = Q,
\end{equation}
and
\begin{equation}
\label{eq41}\dot{\rho}^{X} + 3\frac{\dot{a}}{a}\left(\rho^{X} +
p^{X}_{eff}\right) = \dot{\rho}^{X} + (1 + \omega^{X}_{eff})
\rho^{X}(2n + 1)\frac{\dot{C}}{C} = - Q.
\end{equation}
The quantity $Q$ expresses the interaction between the dark components. Since we are interested in
an energy transfer from the dark energy to dark matter, we consider $Q > 0$. $Q > 0$, ensures that the second
law of thermodynamics is fulfilled (Pavon \& Wang 2009). Here we
emphasize that the continuity Eqs. (\ref{eq40}) and (\ref{eq41})
imply that the interaction term ($Q$) should be  proportional to a
quantity with units of inverse of time i.e $Q\propto \frac{1}{t}$.
Therefore, a first and natural candidate can be the Hubble factor
$H$ multiplied with the energy density. Following Amendola et al (2007) and Gou et al (2007), we consider
\begin{equation}
\label{eq42}Q = H \sigma \rho^{m},
\end{equation}
where $\sigma$ is a coupling constant. Using Eq. (\ref{eq42}) in Eq. (\ref{eq40}) and after integrating, we obtain
\begin{equation}
\label{eq43}\rho^{m} = \rho_{0}C^{-(2n + 1)(1 + \omega^{m} -
\sigma)} = \rho_{0} \l \sinh^{-3(1 + \omega^{m} -\sigma)} {(T)},
\end{equation}
where $\l = \ell_{3}^{-(2n + 1)(1 + \omega^{(m)} -\sigma)}$. \\

By using Eqs. (\ref{eq20}), (\ref{eq21}) and (\ref{eq43}) in Eqs.
(\ref{eq13}) and (\ref{eq16}), we obtain
\begin{equation}
\label{eq44} \rho^{X} = n(n + 2)\frac{\dot{C}^{2}}{C^{2}} -
\frac{\alpha^{2}}{C^{2n}} - \rho_{0}C^{-(2n + 1) (1 + \omega^{m} -
\sigma)},
\end{equation}
and
\begin{equation}
\label{eq45} p^{X} = - \left[2n\frac{\ddot{C}}{C} + n(3n -
2)\frac{\dot{C}^{2}}{C^{2}} - \frac{\alpha^{2}}{C^{2n}} \right] -
\rho_{0}(\omega^{m} - \sigma)C^{-(2n + 1)(1 + \omega^{(m)} -
\sigma)} .
\end{equation}
Using Eq. (\ref{eq26}) in Eqs. (\ref{eq44}) and (\ref{eq45}), we
obtain the values of $\rho^{X}$ and $p^{X}_{eff}$ as
\[
\rho^{X} = \frac{9n(n + 2)}{(2n + 1)^{2}}\coth^{2}(t) -
\alpha^{2}\ell^{-2n}_{3}\sinh^{-\frac{6n} {(2n + 1)}}(t) -
\rho_{0}l_{0}\sinh^{-3(1 + \omega^{m}-\sigma)}{(t)}
\]
\begin{equation}
\label{eq46}= \frac{9n(n + 2)}{(2n +
1)^{2}}\left[1+(1+z)^{2}\right] -
\alpha^{2}\ell^{-2n}_{3}(1+z)^{\frac{6n} {(2n + 1)}} -
\rho_{0}l_{0}(1+z)^{3(1 + \omega^{m}-\sigma)}
\end{equation}
and
\[
p^{X}_{eff}= -\left[\frac{9(n^{2} + n + 1)}{(2n +
1)^{2}}\coth^{2}(t)- \frac{3(n + 1)}{(2n + 1)} \cosh^{2}(t)\right]
- (\omega^{m}-\sigma)\rho_{0}l_{0}\sinh^{-3(1 +
\omega^{m}-\sigma)}(t)-3\xi_{0}H(\rho^{X})^{\tau}
\]
\begin{equation}
\label{eq47}= -\left[\frac{9(n^{2} + n + 1)}{(2n +
1)^{2}}\left[1+(1+z)^{2}\right]- \frac{3(n + 1)}{(2n + 1)}
\left[1+(1+z)^{-2}\right]\right] -
(\omega^{m}-\sigma)\rho_{0}l_{0}(1+z)^{3(1 +
\omega^{m}-\sigma)}-3\xi_{0}H(\rho^{X})^{\tau}.
\end{equation}
respectively.  \\
\begin{figure}[ht]
\begin{minipage}[b]{0.5\linewidth}
\centering
\includegraphics[width=\textwidth]{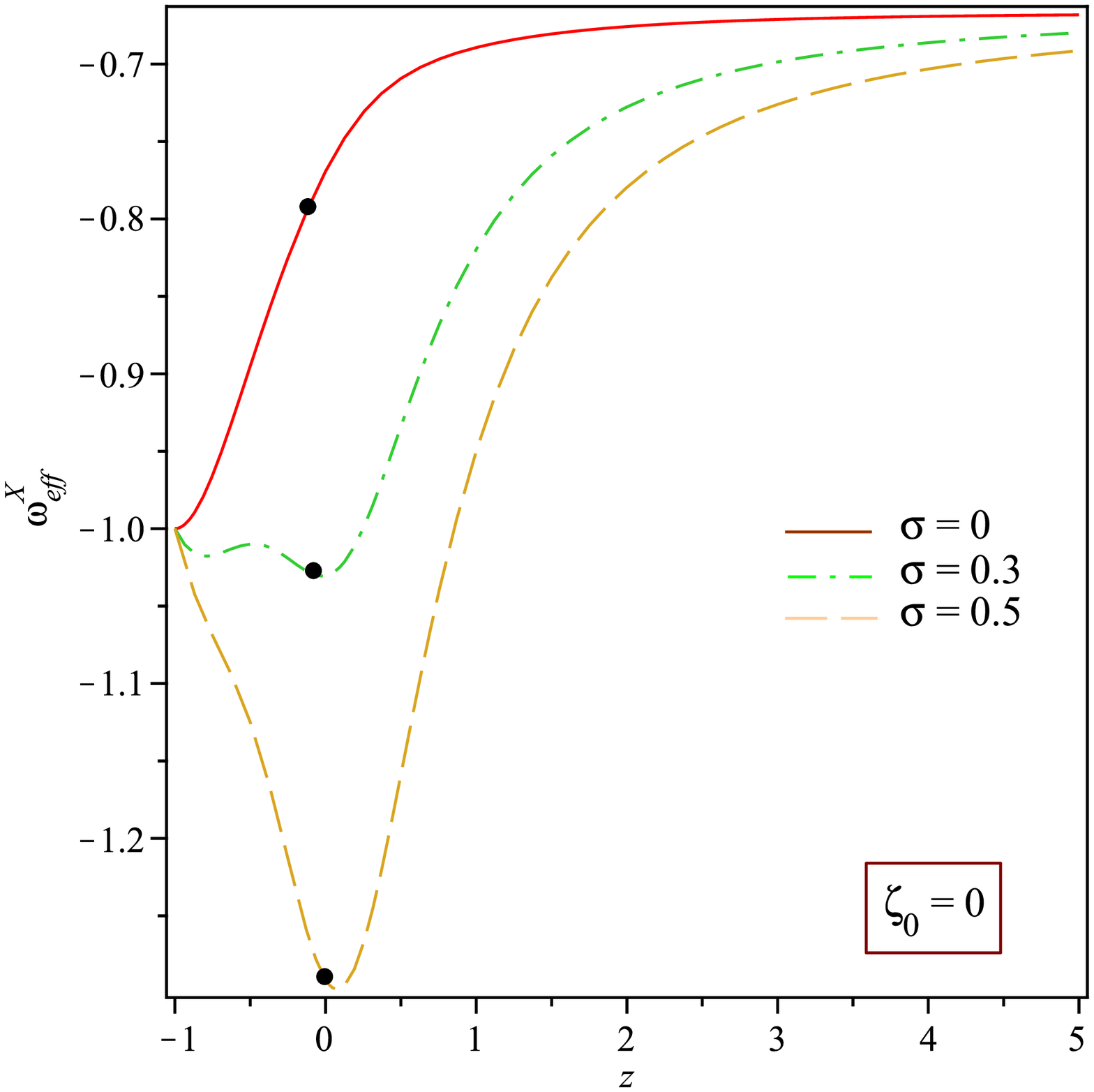} \\
\caption{The EoS parameter $\omega^{X}_{eff}$ versus $z$ for $n =
\beta = \alpha = \ell_{3} = l_{0}=1$, $\Omega_{0}^{m} = 0.3$. The
dots locate the current values of $\omega^{X}_{eff}$. In this
case, we fix $\zeta_{0}=0$ and vary $\sigma$.}
\end{minipage}
\hspace{0.5cm}
\begin{minipage}[b]{0.5\linewidth}
\centering
\includegraphics[width=\textwidth]{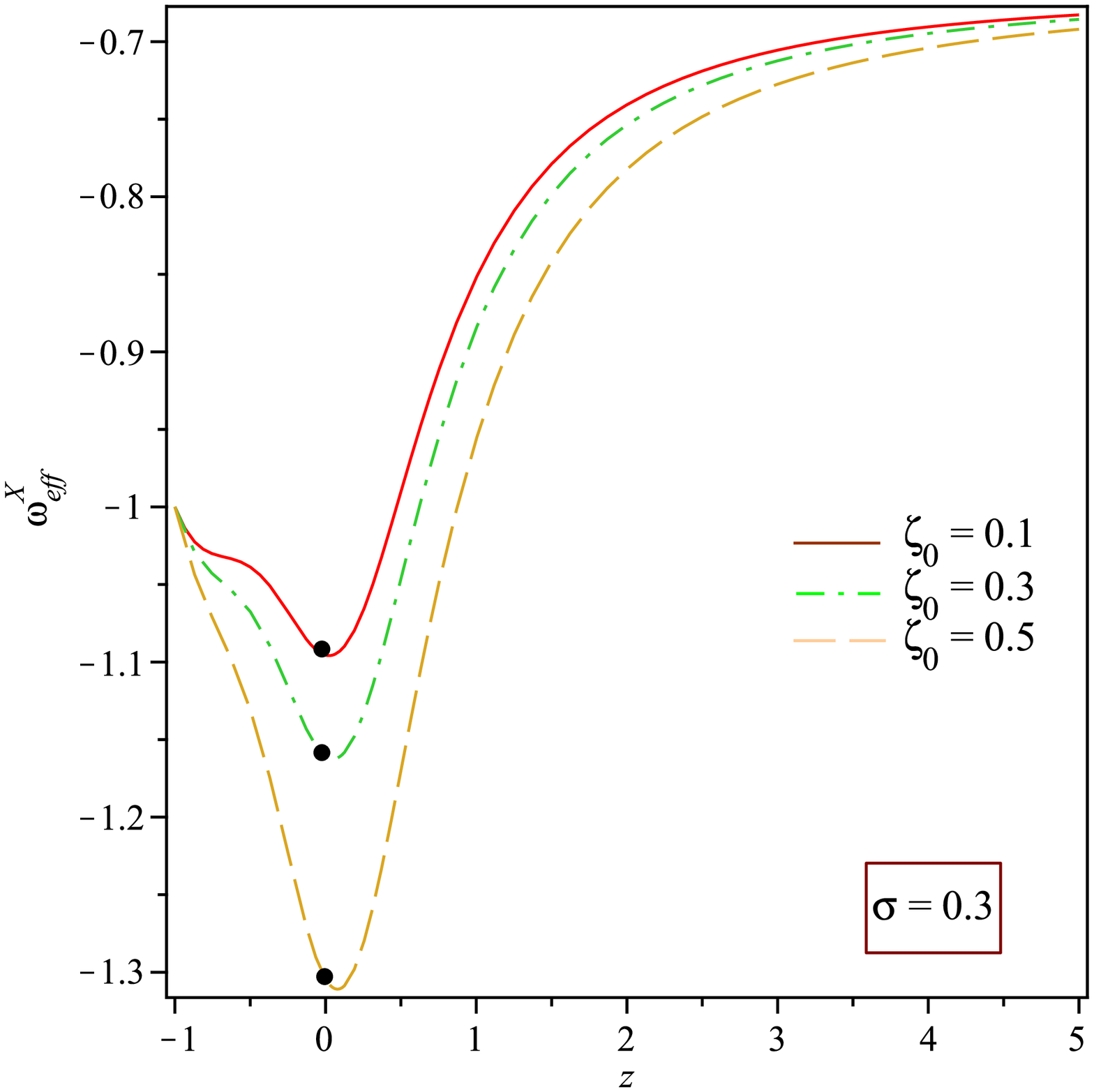}
\caption{The EoS parameter $\omega^{X}_{eff}$ versus $z$ for $n =
\beta = \alpha = \ell_{3} = l_{0}=1$, $\Omega_{0}^{m} = 0.3$. The
dots locate the current values of $\omega^{X}_{eff}$. In this
case, we fix $\sigma=0.3$ and vary $\zeta_{0}$.}
\end{minipage}
\end{figure}
Also the EoS parameter for DE ($\omega^{X}_{eff}$) is obtained as
\[
\omega^{X}_{eff} = - \left[\frac{\frac{9(n^{2} + n + 1)}{(2n +
1)^{2}}\coth^{2}(t)- \frac{3(n + 1)}{(2n + 1)}\cosh^{2}{(t)} +
3\Omega^{m}_{0}l_{0}(\omega^{m}-\sigma)\sinh^{-3(1 -
\omega^{m})}(t)} {\frac{9n(n + 2)}{(2n + 1)^{2}}\coth^{2}(t) -
\alpha^{2}\ell^{-2n}_{3} \sinh^{-\frac{6n}{(2n + 1)}}{(t)} -
3\Omega^{m}_{0}l_{0}\sinh^{-3(1 +
\omega^{m})}(t)}\right]-\zeta_{0}H^{\tau}(\Omega^{X})^{\tau-1}
\]
\begin{equation}
\label{eq48}= - \left[\frac{\frac{9(n^{2} + n + 1)}{(2n +
1)^{2}}\left[1+(1+z)^{2}\right]- \frac{3(n + 1)}{(2n +
1)}\left[1+(1+z)^{-2}\right] +
3\Omega^{m}_{0}l_{0}(\omega^{m}-\sigma)(1+z)^{3(1 - \omega^{m})}}
{\frac{9n(n + 2)}{(2n + 1)^{2}}\left[1+(1+z)^{2}\right] -
\alpha^{2}\ell^{-2n}_{3} (1+z)^{\frac{6n}{(2n + 1)}} -
3\Omega^{m}_{0}l_{0}(1+z)^{3(1 +
\omega^{m})}}\right]-\zeta_{0}H^{\tau}(\Omega^{X})^{\tau-1}.
\end{equation}
The behavior of EoS ($\omega^{X}_{eff}$) parameter for dark energy
in terms of red shift $z$ is shown in Figures. $5, 6$. Again,
since we are interested in the late time and future evolution of
DE, we plot the range of red shift $z$ from $-1$ to $z=5$. Here
the parameter $\omega^{m}$ is taken to be $0$. In Fig. $5$ we fix
the parameter $\zeta_{0}=0$ and vary $\sigma$ as $0$, $0.3$, and
$0.5$ respectively; in Fig. $6$ we fix $\sigma=0.3$ and vary
$\zeta_{0}$ as $0.1$, $0.3$, and $0.5$ respectively. The plots
show that the evolution of $\omega^{X}_{eff}$ depends on the
parameters $\sigma$ and $\zeta_{0}$ apparently. It is clear that
(from Fig. $5$) the interaction alleviate the EoS parameter of DE
to go to darker regions as in non-interacting case (Fig. $1$). But
considering the bulk viscosity in the cosmic fluid, compensates
the effect
of interaction (see Fig. $6$).\\
\begin{figure}[ht]
\begin{minipage}[b]{0.5\linewidth}
\centering
\includegraphics[width=\textwidth]{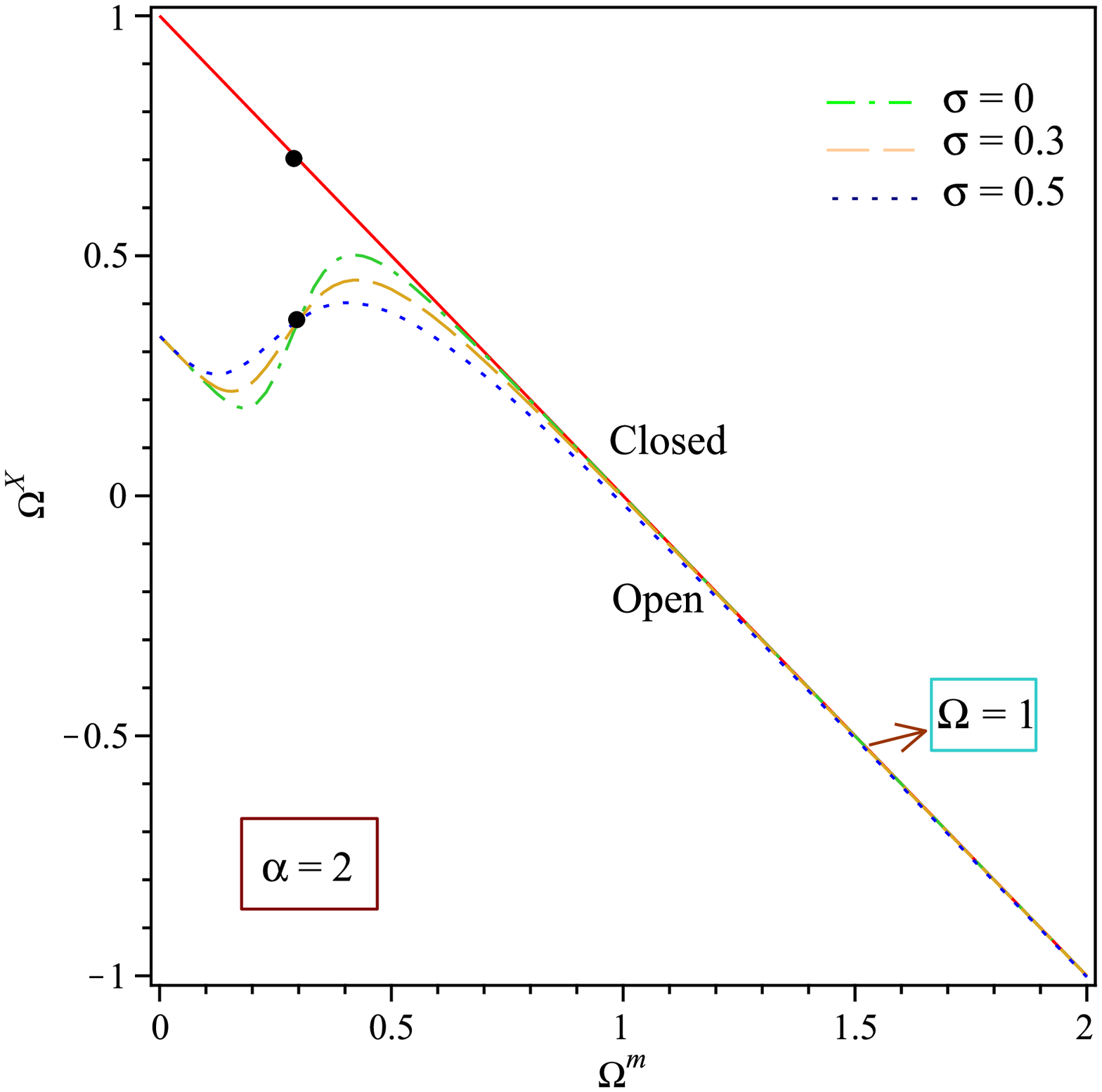} \\
\caption{The plot of $\Omega^{X}$ versus $\Omega^{m}$ for
$n=\ell_{0}=\ell_{3}=1,~\Omega^{m}_{0}=0.3$. The solid line
indicates flat universe ($n=1$,~$\alpha=0$). The dots locate the
current values of $\Omega^{X}$ and $\Omega^{m}$. In this case, we
fix $\alpha=2$ and vary $\sigma$.}
\end{minipage}
\hspace{0.5cm}
\begin{minipage}[b]{0.5\linewidth}
\centering
\includegraphics[width=\textwidth]{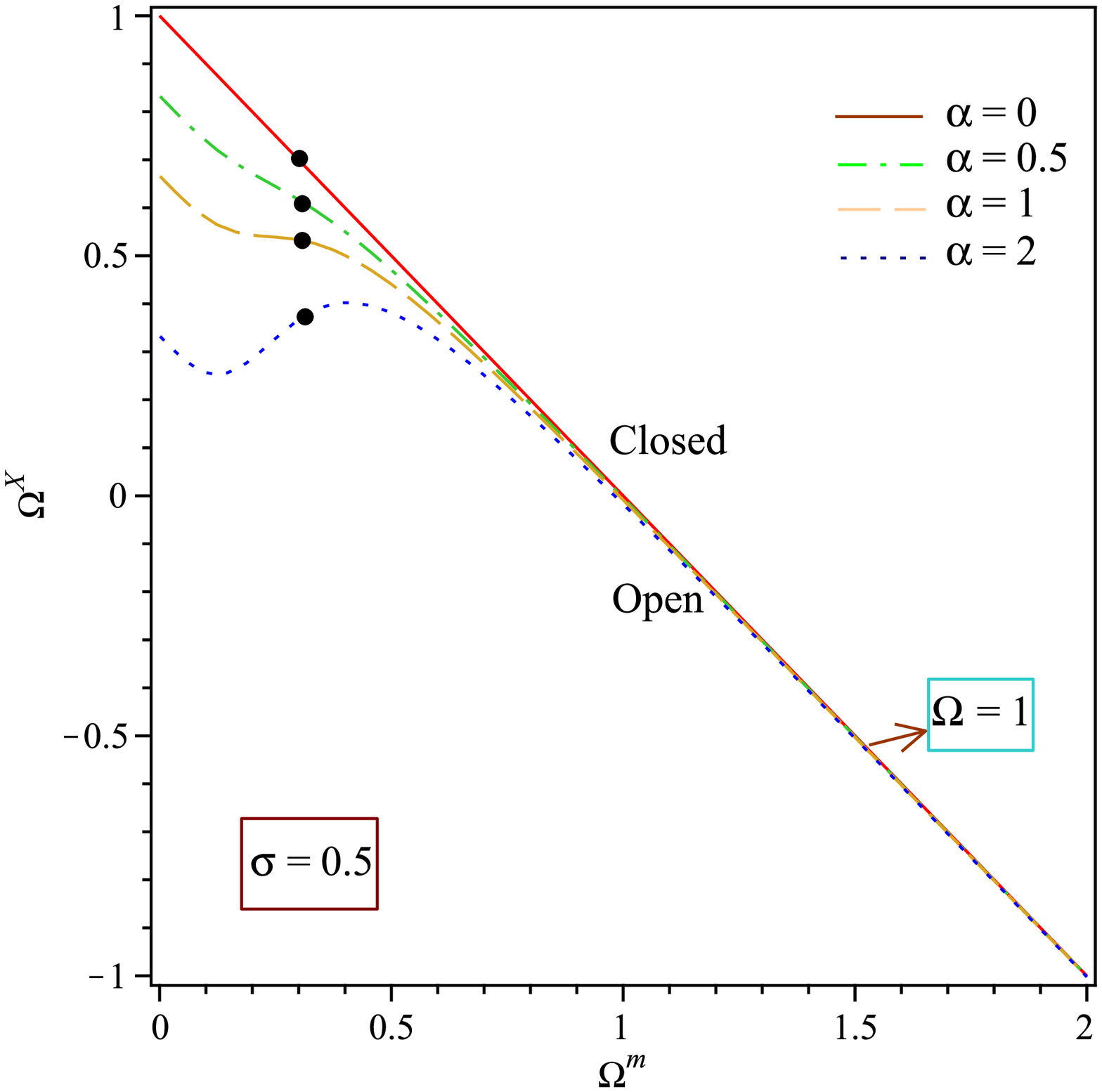}
\caption{The plot of $\Omega^{X}$ versus $\Omega^{m}$ for
$n=\ell_{0}=\ell_{3}=1,~\Omega^{m}_{0}=0.3$. The solid line
indicates flat universe ($n=1$,~$\alpha=0$). The dots locate the
current values of $\Omega^{X}$ and $\Omega^{m}$. In this case, we
fix $\sigma=0.5$ and vary $\alpha$.}
\end{minipage}
\end{figure}
\begin{figure}[htbp]
\centering
\includegraphics[width=10cm,height=10cm,angle=0]{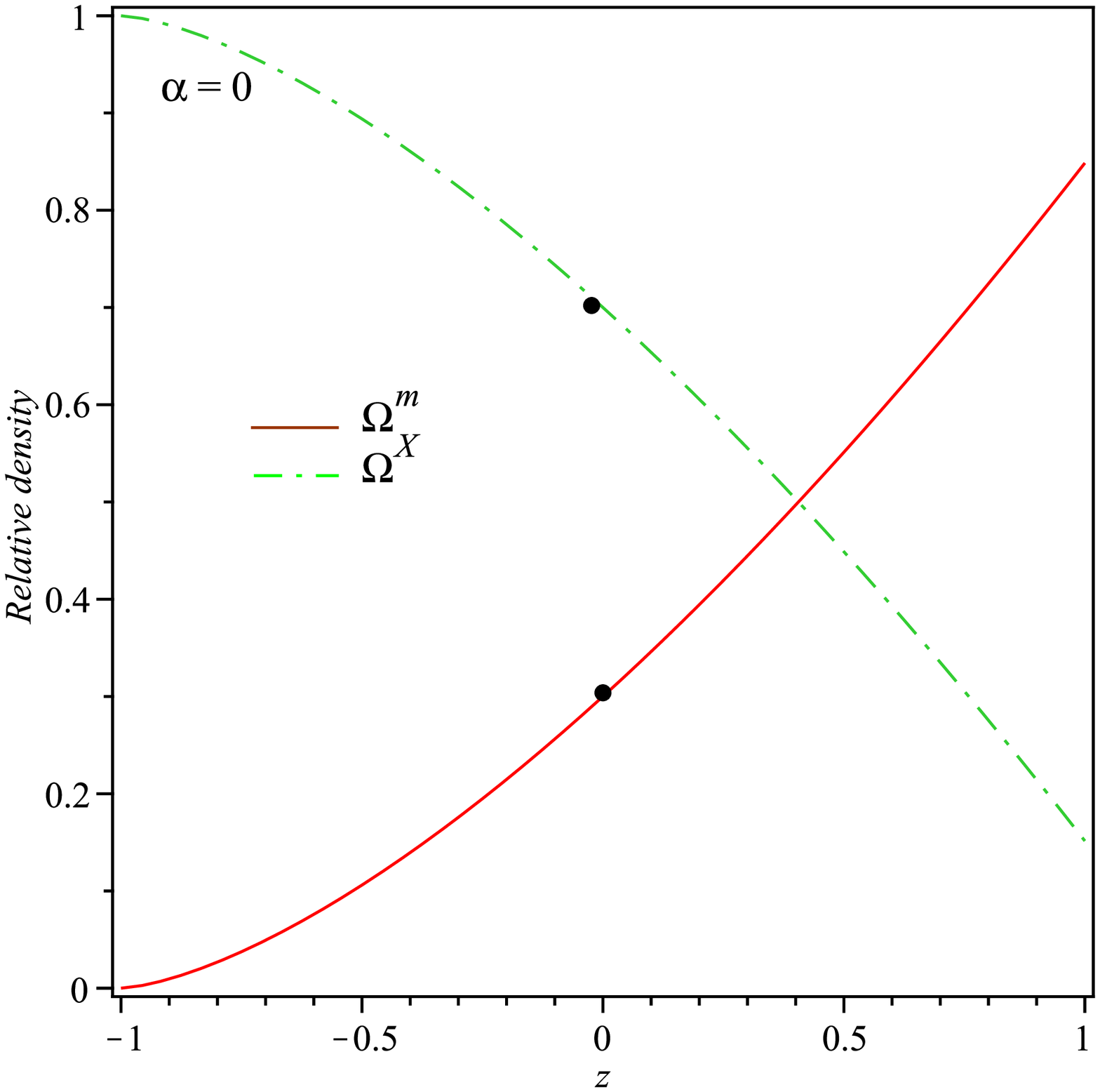}
\caption{The plot of energy $\Omega^{m}$ and $\Omega^{X}$ versus
redshift ($z$) for $\Omega^{m}_{0}=0.3,~
\ell_{0}=n=1,~\sigma=0.5$. The dots locate the current values of
$\Omega^{X}$ and $\Omega^{m}$.}
\end{figure}
The expressions for the matter-energy density $\Omega^{m}$ and dark-energy density $\Omega^{X}$ are given by
\begin{equation}
\label{eq49}\Omega^{m} = \frac{\rho^{m}}{3H^{2}} =
\frac{\rho_{0}l_{0}\sinh^{-3(1 + \omega^{m}-\sigma)}{(t)}}
{3\coth^{2}{(t)}}=\Omega^{m}_{0}l_{0}(1+z)^{3(1 -
\omega^{m}-\sigma)},
\end{equation}
and
\[
\Omega^{X} =  \frac{\rho^{X}}{3H^{2}} =\frac{3n(n + 2)}{(2n +
1)^{2}}- \frac{\alpha^{2}\ell^{-2n}_{3}\sinh^{-\frac{6n}{(2n +
1)}}(t) + \rho_{0}l_{0}\sinh^{-3(1 + \omega^{m}-\sigma)}{(t)}}
{3\coth^{2}{(t)}}
\]
\begin{equation}
\label{eq50} =\frac{3n(n + 2)}{(2n + 1)^{2}}-
\frac{\alpha^{2}\ell^{-2n}_{3}(1+z)^{\frac{6n}{(2n + 1)}}}
{3\left[1+(1+z)^{2}\right]}-\Omega^{m}_{0}l_{0}(1+z)^{3(1 -
\omega^{m}-\sigma)}
\end{equation}
respectively. Adding Eqs. (\ref{eq49}) and (\ref{eq50}), we obtain
total energy ($\Omega$)
\[
\Omega = \Omega^{m} + \Omega^{X} =\frac{9n(n + 2)}{(2n +
1)^{2}}-\frac{\alpha^{2}\ell_{3}^{-2n}\sinh^{-\frac{6n}{2n +
1}}{(t)}}{3\coth^{2}{(t)}}
\]
\begin{equation}
\label{eq51}=\frac{9n(n + 2)}{(2n +
1)^{2}}-\frac{\alpha^{2}\ell_{3}^{-2n}(1+z)^{\frac{6n}{2n +
1}}}{3\left[1+(1+z)^{2}\right]}.
\end{equation}
which is the same as Eq. (\ref{eq38}). Therefore, we observe that in interacting case the density parameter has the
same properties as in non-interacting case.\\

The values of $\Omega^{m}$ and $\Omega^{X}$ which are permitted by
our models in interacting case are shown in Figures. $7, 8$. In
both figures the line $1=\Omega^{m}$ + $\Omega^{X}$  indicates a
flat universe separating open from closed universes. In Fig. $7$
we fix the parameter $\alpha=2$ and vary $\sigma$ as $0$, $0.3$,
and $0.5$ respectively; in Fig. $8$ we fix $\sigma=0.5$ and vary
$\alpha$ as $0$, $0.5$, $1$, and $2$ respectively. The plots show
that the evolution of $\Omega^{X}$ versus $\Omega^{m}$ depends on
the parameters $\sigma$ and $\alpha$ apparently. Fig. $9$ depicts
the evolution of the relative densities. From this figure we
observe that the interaction parameter $\sigma$ brings impact on
the evolution of the densities depending to its value.
\section{Statefinder Diagnostic}
Since there are many models suggested in order to
describe the current cosmic acceleration, it is very important to
find a way to discriminating between the various contenders in a
model-independent manner. For this purpose, Sahni et al (2003) have
introduced a new cosmological diagnostic pair $\{s, r\}$ called
the statefinder. The parameters$s$ and $r$ are dimensionless and
only depend on the scale factor $a$, therefore $\{s, r\}$ is a
geometrical diagnostic. They were defined as
\begin{figure}[ht]
\begin{minipage}[b]{0.5\linewidth}
\centering
\includegraphics[width=\textwidth]{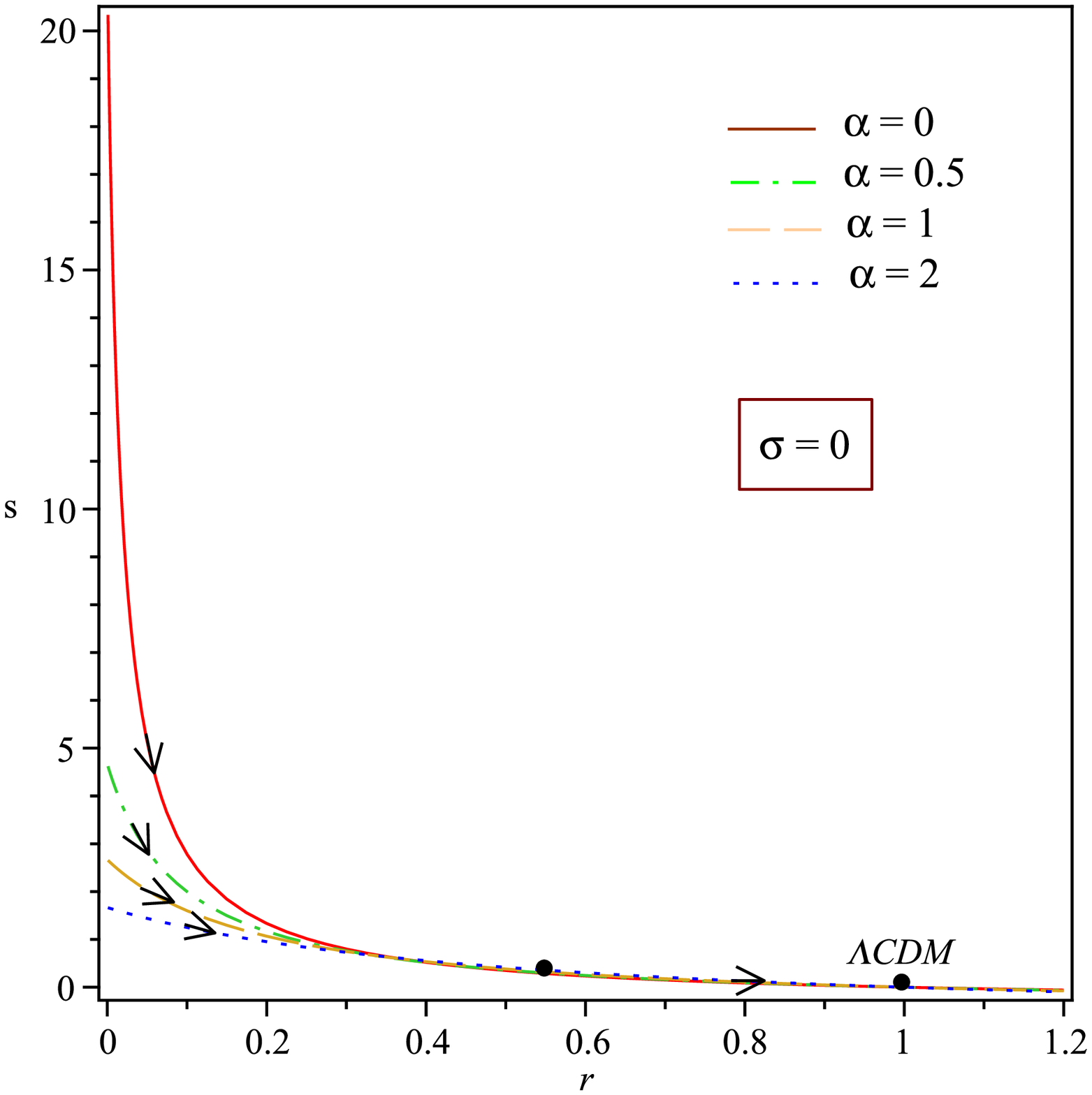} \\
\caption{$s-r$ evolution diagram. The dots locate the current
values of the statefinder pair $\{s,r\}$. In this case, we fix
$\sigma=0$ (non-interacting case) and vary $\alpha$.}
\end{minipage}
\hspace{0.5cm}
\begin{minipage}[b]{0.5\linewidth}
\centering
\includegraphics[width=\textwidth]{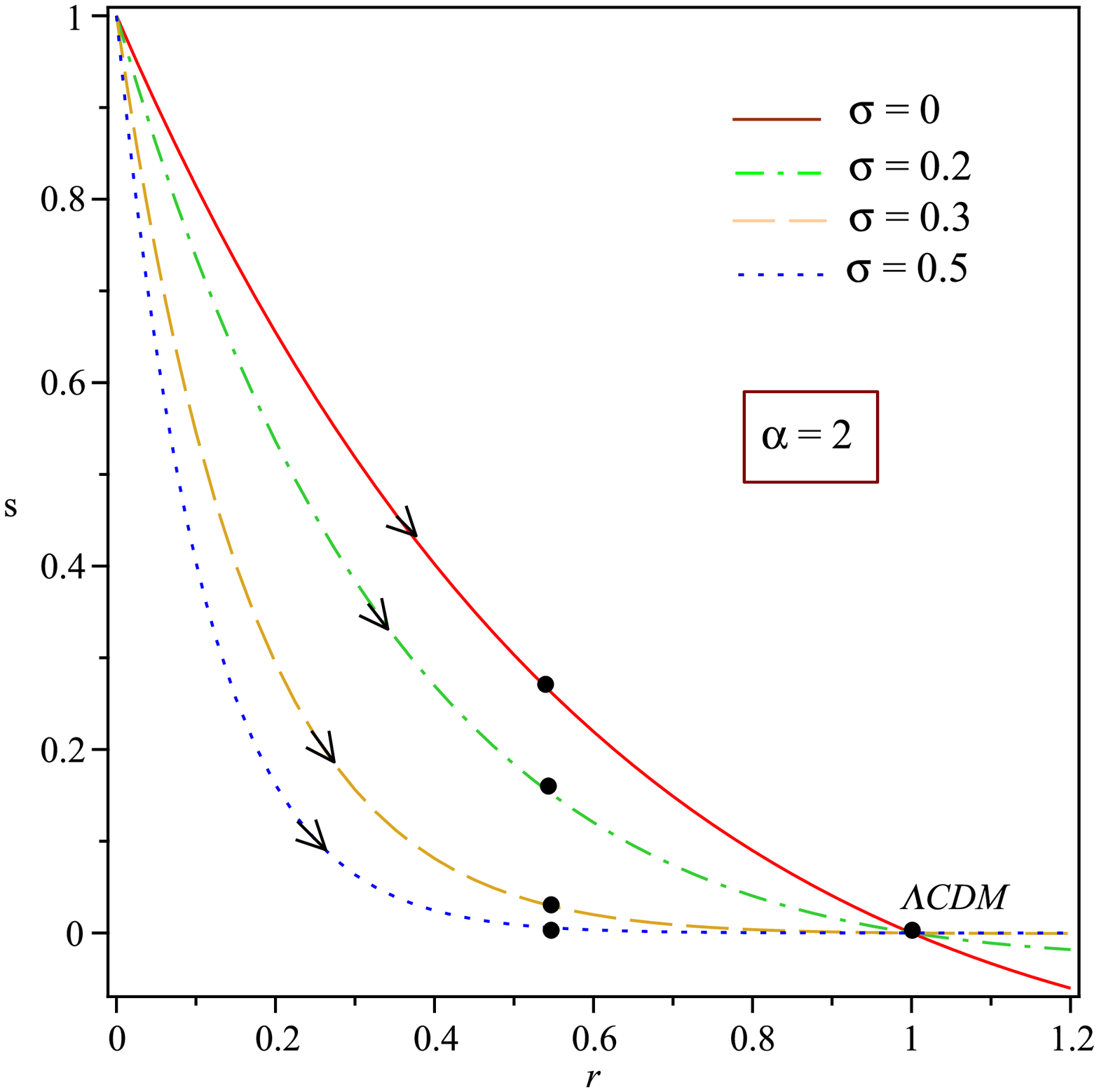}
\caption{$s-r$ evolution diagram. The dots locate the current
values of the statefinder pair $\{s,r\}$. In this case, we fix
$\alpha=2$ (interacting case) and vary $\sigma$.}
\end{minipage}
\end{figure}
\begin{equation}
\label{eq52}
r\equiv\frac{\dot{\ddot{a}}}{aH^{3}},~~~~~s\equiv\frac{r-\Omega}{3(q-\frac{\Omega}{2})}.
\end{equation}
Here the formalism of Sahni and coworkers is extended to permit
curved universe models. Using these parameters one can
differentiate between different forms of dark energy. For example,
although the quintessence, phantom and Chaplygin gas models tend
to approach the $\Lambda$CDM fixed point ($\{s,r\}_{\Lambda
CDM}=\{0,1\}$), for quintessence and phantom models the
trajectories lie in the region $s>0,~r<1$ whereas for
Chaplygin gas models trajectories lie in region $s<0,~r>1$.\\
In general, the statefinder parameters are given by
\begin{equation}
\label{eq53}
r=\Omega^{m}+\frac{9\omega^{X}}{2}\Omega^{X}(1+\omega^{X})-\frac{3}{2}\Omega^{X}\frac{\dot{\omega}^{X}}{H},
\end{equation}
\begin{equation}
\label{eq54}
s=1+\omega^{X}-\frac{1}{3}\frac{\dot{\omega}^{X}}{\omega^{X}H}.
\end{equation}
Since we have the analytical expression of $\omega^{X}_{eff}$ in
both non-interacting and interacting cases we can easily obtain
$\frac{\dot{\omega}^{X}_{eff}}{H}$. Thus, we can calculate the
statefinder parameters in this scenario.\\

The evolution of the ststefinder pair $\{s,r\}$ is shown in
Figures. $10, 11$. In Fig. $10$ we fix the parameter $\sigma=0$
and vary $\alpha$ as $0$, $0.5$, $1$, and $2$ respectively; in
Fig. $11$ we fix $\alpha=2$ and vary $\sigma$ as $0$, $0.2$,
$0.3$, and $0.5$ respectively. The filled circles show the current
values of statefinder pair $\{s,r\}$ for different dark energy
models. Here, we observe that the interaction parameter $\sigma$
makes the model evolve along different trajectories on the $s-r$
plane.
\section{Concluding Remarks}
In this paper we studied dark energy in the scope of anisotropic Bianchi type III space-time. We considered two cases (i) when DE and DM do not interact with each other and (ii) when there is an interaction between these two dark components. In non-interacting as well as weak interacting ($\sigma\sim 0$) cases we observed that in absence of viscosity, dark energy EoS parameter dose not cross the phantom divided line (PDL) and hence always vary in quintessence region. However, in both cases when dark energy is considered to be viscous rather than perfect, it's EoS parameter could cross the PDL depending on the values of coupling constant $\sigma$ and bulk viscosity coefficient $\zeta_{0}$. But in this case although the dark energy EoS parameter could cross PDL and vary in phantom region ultimately tends to the cosmological constant region $\omega^{de}=-1$. This special behavior of the EoS parameter is because of our choose of bulk viscosity which is a decreasing function of time (redshift) in expanding universe. It has also been shown that in both cases according to the $\Omega^{X}$-$\Omega^{m}$ phase diagram (see figs. 3, 7, 8), deviation from flat universe ($\Omega=1$) only depends on the geometric parameter $\alpha$ not to the interaction parameter $\sigma$.

\section*{Acknowledgments}
Author would like to thank Mahshahr branch of Islamic Azad University for providing facility and support where
this work was carried out.

\end{document}